\documentstyle[12pt,a41,axodraw]{article}

\begin{document}

\begin{flushleft}
INLO-PUB 05/00\hfill {\tt hep-ph/xxyyzz}\\
March 2000\\
\end{flushleft}
\vspace*{\fill}
\begin{center}
{\LARGE\bf Higher Order Corrections in Perturbative \footnote{Talk presented 
at WHEPP "Sixth Workshop in High
Energy Physics Phenomenology", Chennai (Madras) 600 113, India, January 3-15,
2000.}}
\vspace{2mm}

{\LARGE\bf Quantum Chromo Dynamics}

\vspace*{20mm}
\large
{W.L. van Neerven \footnote{Work supported by the
EC network `QCD and Particle Structure'  under contract No.~FMRX--CT98--0194.}}
\\

\vspace{2em}

\normalsize
{\it Instituut-Lorentz, Universiteit Leiden, PO Box 9506, 
2300 RA Leiden, The Netherlands.} 

\vspace{2em}
\end{center}
\vspace*{\fill}
\begin{abstract}
\noindent
We present some techniques which have been developed recently
or in the recent past to compute Feynman graphs beyond one-loop order.
These techniques are useful to compute the three-loop splitting
functions in QCD and to obtain the complete second order QED corrections to
Bhabha scattering.
\end{abstract}
\vspace*{1cm}
\section{Introduction}
Since the discovery of asymptotic freedom in non-Abelian gauge field theories,
like Quantum Chromo Dynamics (QCD), many perturbative calculations have been 
performed to hadron-hadron and semi-leptonic processes. The most reactions
are computed up to next-to-leading order (NLO) only, except for some 
semi-leptonic processes. The latter is due to the simplicity of the Born 
approximation where the basic process is given by the interaction of the 
intermediate vector boson with a quark. Higher order corrections to the other 
reactions are still missing because of the very complicated Feynman and phase 
space integrals which arise in the calculations. Here we will discuss some new
methods which maybe will enable us to compute the QCD and also QED corrections
beyond the NLO level.
\section{Asymptotic Expansions}
In this section we concentrate on the computation of the coefficient
functions and anomalous dimensions which show up in deep inelastic 
lepton-hadron scattering. The relevant quantity in this process is the
structure function defined by
\begin{eqnarray}
\label{eqn2.1}
F(x,Q^2)= \frac{1}{4\pi} \int d^4z \, e^{iq.z} \langle p \mid
[J(z),J(0)] \mid p \rangle \quad q^2=-Q^2 < 0 \quad 
x=\frac{Q^2}{2\nu}\,.
\end{eqnarray}
In the commutator above $J(z)$ represents the electro-weak current where
we have suppressed Lorentz indices for convenience. In the Bjorken limit
i.e. $Q^2 \rightarrow \infty$ and $x=fixed$ 
the light cone region dominates the integrand so that one can make an Operator 
Product Expansion (OPE) 
\begin{eqnarray}
\label{eqn2.2}
 [J(z),J(0)]
 {\raisebox{-2 mm}{$\,\stackrel{=}{{\scriptstyle z^2 \sim 0 }}\, $} }
 \sum_{N} C^N(z^2 \mu^2) O^N (\mu^2,0)\,.
\end{eqnarray}
This expansion also holds for the time ordered product corresponding to
forward Compton scattering
\begin{eqnarray}
\label{eqn2.3}
T(x,Q^2)&=& \frac{i}{4\pi^2} \int d^4z \, e^{iq.z} \langle p \mid
[J(z),J(0)] \mid p \rangle\,, 
\nonumber\\[2ex]
 {\rm Im}\, T(x,Q^2) &=& \pi \, F(x,Q^2) \,.
\end{eqnarray}
Let us assume that we can write an unsubtracted dispersion relation
in the variable $\nu = p\cdot q$  
\begin{eqnarray}
\label{eqn2.4}
T(\nu,Q^2)= \int_{Q^2/2}^{\infty} d\nu' \frac{F(\nu',Q^2)}{\nu' -\nu}\,,
\end{eqnarray}
so that after substitution of the variables
$\nu= \frac{Q^2}{2x}$, $\nu'= \frac{Q^2}{2x'}$ we can derive
the following relation
\begin{eqnarray}
\label{eqn2.5}
T(\nu,Q^2)&=&x \int_0^1 \frac{dx'}{x'} \frac{F(x',Q^2)}{x -x'} 
\nonumber\\[2ex]
&=& \sum_{N=0}^{\infty} \left (\frac{2p\cdot q}{Q^2} \right )^N
\int_0^1 dx' \, x'^{N-1} F(x',Q^2)
\nonumber\\[2ex]
&=& \sum_{N=0}^{\infty} \left (\frac{2p\cdot q}{Q^2} \right )^N
A^{(N)}(\mu^2) {\cal C}^{(N)} \left (\frac{Q^2} {\mu^2} \right )\,,
\end{eqnarray}
where the operator matrix element and the coefficient function are given by
\begin{eqnarray}
\label{eqn2.6}
A^{(N)}(\mu^2)=\langle p \mid O^N(\mu^2,0) \mid p \rangle \,,
\end{eqnarray}
and
\begin{eqnarray}
\label{eqn2.7}
{\cal C}^{(N)}\left(\frac{Q^2}{\mu^2}\right)=\int d^4 z \,e^{iq.z}
C^N(z^2 \mu^2) \,,
\end{eqnarray}
respectively. Both quantities satisfy a renormalization group equation
\begin{eqnarray}
\label{eqn2.8}
&& \left [ \mu \frac{\partial}{\partial \mu} + \beta (g) \frac{\partial}
{\partial g} + \gamma^{(N)}(g) \right ] A^{(N)}(\mu^2)=0\,,
\nonumber\\[2ex]
&& \left [ \mu \frac{\partial}{\partial \mu} + \beta (g) \frac{\partial}
{\partial g} - \gamma^{(N)}(g) \right ] {\cal C}^{(N)}\left(\frac{Q^2}{\mu^2}
\right) =0\,.
\end{eqnarray}
In the equations above the quantities $\beta(g)$ and $\gamma^{(N)}(g)$
represent the beta-function and the anomalous dimension respectively. 
The latter determine the $Q^2$-evolution of the structure function $F(x,Q^2)$
which can be measured in experiment and provides us with one of the tests
of perturbative QCD. If one computes the coefficient functions in the
conventional way one encounters phase space integrals and loop integrals
(see e.g. \cite{zn}). However if one wants to compute the coefficient
function beyond order $g^4$ it is more convenient to try another method
which is explained in \cite{move}. Taking $\phi_6^3$-theory as an example
one can expand the propagator in $T(x,Q^2)$ (\ref{eqn2.3}) given by
$1/(k-p)^2$ as follows
\begin{eqnarray}
\label{eqn2.9}
T(x,Q^2)\equiv T&=&(-ig)^2 \int \frac{d^nk}{(2\pi)^n} 
\frac{i^3}{(k^2)^2(k-p)^2(k+q)^2}
\nonumber\\[2ex]
&=&-ig^2 \sum_{N=0}^{\infty} \int \frac{d^nk}{(2\pi)^n} \frac{(2k \cdot p)^N}
{(k^2)^{3+N}(k+q)^2} \,,
\end{eqnarray}
\begin{figure}
\begin{center}
  \begin{picture}(260,100)(0,0)
  \Photon(0,0)(20,20){2}{5}
  \DashArrowLine(20,20)(20,80){3}
  \ArrowLine(0,100)(20,80)
  \ArrowLine(20,80)(60,80)
  \DashArrowLine(60,80)(100,80){3}
  \DashArrowLine(100,80)(100,20){3}
  \ArrowLine(100,20)(120,0)
  \ArrowLine(60,80)(60,20)
  \DashArrowLine(60,20)(20,20){3}
  \ArrowLine(100,20)(60,20)
  \Photon(100,80)(120,100){2}{5}
  \Text(15,100)[t]{$p$}
  \Text(105,100)[t]{$q$}
  \Text(45,95)[t]{$p+k_1$}
  \Text(85,95)[t]{$q+k_2$}
  \Text(10,55)[t]{$k_1$}
  \Text(110,55)[t]{$k_2$}
  \Text(45,15)[t]{$k_1-q$}
  \Text(85,15)[t]{$k_2-p$}
  \Text(15,10)[t]{$q$}
  \Text(105,10)[t]{$p$}
  \Text(60,50)[t]{$k_1-k_2-q+p$}
  \Text(130,50)[t]{$\rightarrow$}
  \Photon(140,50)(165,50){2}{5}
  \Photon(235,50)(260,50){2}{5}
  \DashArrowArcn(200,50)(35,180,90){3}
  \DashArrowArcn(200,50)(35,90,0){3}
  \DashArrowArcn(200,50)(35,270,180){3}
  \DashArrowArcn(200,50)(35,360,270){3}
  \DashArrowLine(200,85)(200,15){3}
  \Text(168,85)[t]{$k_1$}
  \Text(240,85)[t]{$k_2+q$}
  \Text(165,25)[t]{$k_1-q$}
  \Text(150,45)[t]{$q$}
  \Text(250,45)[t]{$q$}
  \Text(235,25)[t]{$k_2$}
  \Text(205,65)[t]{$k_1-k_2-q$}
  \end{picture}
  \caption[]{Reduction of a two-loop Compton graph into a self energy
diagram.}
  \label{fig1}
\end{center}
\end{figure}
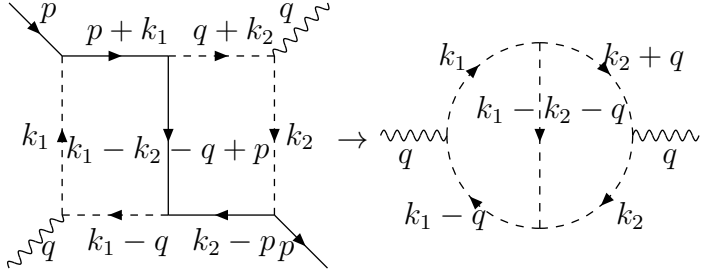
where we have used n-dimensional regularization to regularize the collinear (C)
and ultraviolet (UV) singularities. The asymptotic expansion of the propagator 
converts the Compton amplitude into a self energy type of integral. The
latter is much easier to compute than expressions for box- or triangle graphs.
Another feature is that this asymptotic expansion transforms the C-divergence 
in $T$ at $n=6$ into an UV singularity appearing in the expression for
the self energy integral. Therefore the operator matrix element in Eq. 
(\ref{eqn2.5}), containing the collinear divergence denoted by 
$1/\varepsilon_C$, will be replaced by the corresponding operator 
renormalization constant $Z_{O^N}$ 
\begin{eqnarray}
\label{eqn2.10}
Q^2 T&=&\sum_{N=0}^{\infty} \left (\frac{2p\cdot q}{Q^2} \right )^N
A^{(N)}\left (\frac{1}{\varepsilon_C},\mu^2 \right ) {\cal C}^{(N)}
\left (\frac{Q^2}{\mu^2} \right )
\nonumber\\[2ex]
&& \rightarrow \quad \sum_{N=0}^{\infty} \left ( \frac{2p\cdot q}{Q^2}
\right )^N Z_{O^N} \left (\frac{1}{\varepsilon_{UV}},\mu^2 \right )
{\cal C}^{(N)} \left (\frac{Q^2}{\mu^2}\right )\,,
\end{eqnarray}
where $Z_{O^N}$ depends on the UV singularity denoted by $1/\varepsilon_{UV}$.
This transformation leaves the coefficient function unaltered.
A straightforward calculation yields
\begin{eqnarray}
\label{eqn2.11}
Q^2 T&=& g^2 \frac{\pi^{n/2}}{(2\pi)^n}\frac{\Gamma(n/2-1)\Gamma(n/2-3)}
{\Gamma(n-4)}
\sum_{N=0}^{\infty}
\left (\frac{2p\cdot q}{Q^2} \right )^N \frac{\Gamma(4+N-n/2)}{\Gamma(3+N)}
\left (\frac{Q^2}{\mu^2} \right )^{n/2-3}\,.
\end{eqnarray}
One can expand the gamma-functions above around $\varepsilon=n-6$ and obtain
\begin{eqnarray}
\label{eqn2.12}
Q^2 T&=& \frac{g^2}{64\pi^3} \sum_{N=0}^{\infty}
\left (\frac{2p\cdot q}{Q^2} \right )^N \Big [\frac{1}{N+1}-\frac{1}{N+2}\Big ]
\Big [\frac{2}{\varepsilon} +\gamma_E - \ln 4\pi 
\nonumber\\[2ex]
&& + \ln \frac {Q^2}{\mu^2} - 1 - \sum_{k=1}^N \frac{1}{k} \Big ]\,.
\end{eqnarray}
In order $g^2$ Eq. (\ref{eqn2.10}) can be expressed as
\begin{eqnarray}
\label{eqn2.13}
Q^2 T= \frac{g^2}{64\pi^3}\sum_{N=0}^{\infty} \left [ \frac{\gamma_O^{(N)}}{2}
\left \{ \frac{2}{\varepsilon}+\gamma_E - \ln 4\pi +\ln \frac {Q^2}{\mu^2}
\right \} + c_q^{(N)} \right ]\,,
\end{eqnarray}
so that one can read $\gamma_O^{(N)}$ and $c_q^{(N)}$. If the method
is extended up to order $g^4$ one encounters graphs where at least three
propagators, carrying the momentum $p$ (see the solid line in Fig. 
(\ref{fig1})), have to be expanded
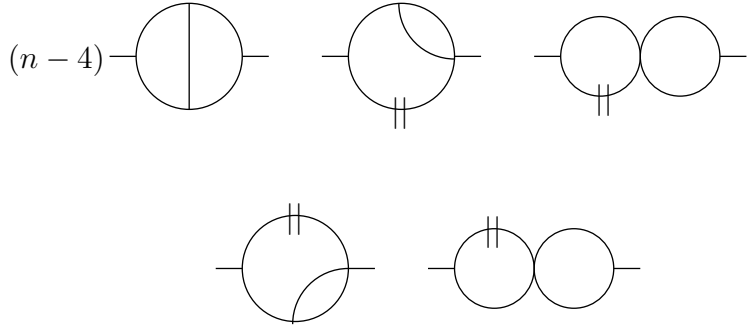
\begin{figure}
\begin{center}
  \begin{picture}(240,160)(0,0)
  \Line(20,120)(30,120)
  \Line(70,120)(80,120)
  \Line(50,140)(50,100)
  \CArc(50,120)(20,0,360)

  \Line(100,120)(110,120)
  \Line(150,120)(160,120)
  \CArc(150,140)(21,180,270)
  \CArc(130,120)(20,0,360)

  \Line(180,120)(190,120)
  \Line(250,120)(260,120)
  \CArc(205,120)(15,0,360)
  \CArc(235,120)(15,0,360)

  \Line(60,40)(70,40)
  \Line(110,40)(120,40)
  \CArc(90,40)(20,0,360)
  \CArc(110,19)(21,90,180)

  \Line(140,40)(150,40)
  \Line(210,40)(220,40)
  \CArc(165,40)(15,0,360)
  \CArc(195,40)(15,0,360)

  \Text(0,125)[t]{$(n-4)$}
  \Text(130,105)[t]{$||$}
  \Text(207,110)[t]{$||$}

  \Text(90,65)[t]{$||$}
  \Text(165,60)[t]{$||$}
  \end{picture}
  \caption[]{Diagrammatic representation of Eq. (\ref{eqn2.16}). The symbol
$||$ means that the corresponding propagator appears twice.}
  \label{fig2}
\end{center}
\end{figure}
This procedure leads to triple sums which are very difficult to unravel
\begin{eqnarray}
\label{eqn2.14}
\sum_i \sum_j \sum_l \left (\frac{2p\cdot k_1}{k_1^2} \right )^i
\left (\frac{2p\cdot k_2}{k_2^2} \right )^j
\left (\frac{2p\cdot (k_1-k_2-q)}{(k_1-k_2-q)^2} \right )^j\,.
\end{eqnarray}
To avoid this one has to remove one or more propagators, carrying the momentum
$p$, before doing the
expansion. First one can adopt the method
\begin{enumerate}
\item{Integration by Parts}\\
This trick has been successfully applied to compute self energy graphs 
\cite{chtk}.  An example is
\begin{eqnarray}
\label{eqn2.15}
I=\int \frac{d^nk_1}{(2\pi)^n} \int \frac{d^nk_2}{(2\pi)^n}
\frac{1}{k_1^2k_2^2(k_1+q)^2(k_2+q)^2(k_1-k_2)^2}\,.
\end{eqnarray}
The integral above appears in the partial integration w.r.t. the momentum $k_1$
of the expression below (see Fig. (\ref{fig2}))
\begin{eqnarray}
\label{eqn2.16}
0=\int \frac{d^nk_1}{(2\pi)^n} \int \frac{d^nk_2}{(2\pi)^n}
\frac{\partial}{\partial k_{1,\mu}}
\frac{(k_1-k_2)_{\mu}}{k_1^2k_2^2(k_1+q)^2(k_2+q)^2(k_1-k_2)^2}\,.
\end{eqnarray}
Hence $I$ in Eq. (\ref{eqn2.15}) can be reduced to a set of integrals
each containing one propagator less than in the original expression.
\item{Recursion Relations (Difference Equations)}\\
Another method is developed in \cite{move}. Using the method of "Difference
Equations" one is able to knock out one propagator (see the example in
Fig. (\ref{fig3})). One can show that all graphs can be reduced to two
building blocks only (see Fig. (\ref{fig4})). The latter contain one
propagator carrying the momentum $p$ so that one only obtains a single sum
which is easy to perform. At this moment it is not clear whether the method 
can be extended beyond order $g^4$ (two-loop).
\end{enumerate}
\begin{figure}
\begin{center}
  \begin{picture}(285,70)(0,0)
  \Line(0,65)(5,55)
  \Line(5,55)(35,55)
  \DashLine(35,55)(65,55){3}
  \Photon(65,55)(70,65){2}{3}
  \DashLine(5,55)(5,15){3}
  \Photon(0,5)(5,15){2}{3}
  \DashLine(5,15)(35,15){3}
  \DashLine(65,55)(65,15){3}
  \Line(35,55)(35,15)
  \Line(35,15)(65,15)
  \Line(65,15)(70,5)
  \Text(90,40)[t]{$=a(N)\times$}
  \Text(0,60)[t]{$p$}
  \Text(0,20)[t]{$q$}
  \Text(73,60)[t]{$q$}
  \Text(73,18)[t]{$p$}

  \Line(105,65)(110,55)
  \Line(110,55)(140,55)
  \DashLine(140,55)(170,55){3}
  \Photon(170,55)(175,65){2}{3}
  \DashLine(110,55)(110,15){3}
  \Photon(105,5)(110,15){2}{3}
  \DashLine(110,15)(140,15){3}
  \DashLine(170,55)(170,15){3}
  \Line(140,55)(140,15)
  \Line(140,15)(170,15)
  \Line(170,15)(175,5)
  \Text(194,40)[t]{$+b(N)\times$}

  \Line(210,65)(215,55)
  \Line(215,55)(245,55)
  \DashLine(245,55)(275,55){3}
  \Photon(275,55)(285,65){2}{3}
  \DashLine(215,55)(215,15){3}
  \Photon(210,5)(215,15){2}{3}
  \DashLine(215,15)(245,15){3}
  \Line(245,55)(245,15)
  \DashLine(245,15)(275,55){3}
  \end{picture}
  \caption[]{Difference Equations.}
  \label{fig3}
\end{center}
\end{figure}
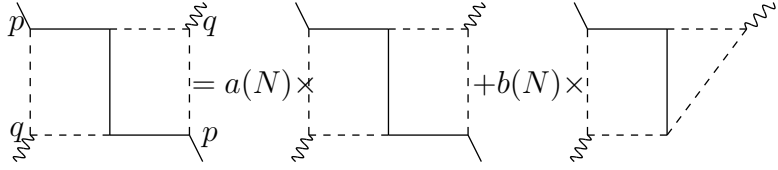

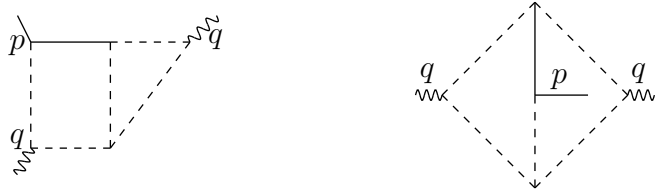
\begin{figure}
\begin{center}
  \begin{picture}(240,70)(0,0)
  \Line(0,65)(5,55)
  \Line(5,55)(35,55)
  \DashLine(35,55)(65,55){3}
  \Photon(65,55)(75,65){2}{3}
  \DashLine(5,55)(5,15){3}
  \Photon(0,5)(5,15){2}{3}
  \DashLine(5,15)(35,15){3}
  \DashLine(35,55)(35,15){3}
  \DashLine(35,15)(65,55){3}

  \Photon(150,35)(160,35){2}{3}
  \Photon(230,35)(240,35){2}{3}
  \DashLine(160,35)(195,70){3}
  \DashLine(160,35)(195,0){3}
  \DashLine(195,70)(230,35){3}
  \DashLine(195,0)(230,35){3}
  \DashLine(195,0)(195,35){3}
  \Line(195,70)(195,35)
  \Line(195,35)(215,35)
  \Text(0,58)[t]{$p$}
  \Text(0,22)[t]{$q$}
  \Text(75,60)[t]{$q$}
  \Text(205,45)[t]{$p$}
  \Text(155,47)[t]{$q$}
  \Text(235,47)[t]{$q$}
  \end{picture}
  \caption[]{Building Blocks.}
  \label{fig4}
\end{center}
\end{figure}
In higher order it is only possible to
obtain the integral over the structure function in Eq. (\ref{eqn2.5}) for
finite moments $N$ (see \cite{lari}). This is achieved by
\begin{eqnarray}
\label{eqn2.17}
\frac{\partial^N T}{\partial p_{\mu_1} \cdots \partial p_{\mu_N}} 
\mid_{p_{\mu_i}=0} \,\, =2^N\,N! \frac{q_{\mu_1} \cdots q_{\mu_N}}{(Q^2)^N} 
Z_{O^N}\left (\frac{1}{\varepsilon_{UV}},\mu^2 \right )
{\cal C}^{(N)} \left (\frac{Q^2}{\mu^2} \right )
\end{eqnarray}
In our example Eq. (\ref{eqn2.9}) this procedure leads to the result
\begin{eqnarray}
\label{eqn2.18}
\frac{\partial^N T}{\partial p_{\mu_1} \cdots \partial p_{\mu_N}} 
\mid_{p_{\mu_i}=0}\,\,
=2^N\,N! (-ig)^2 \int \frac{d^n k}{(2\pi)^n} \frac{ k_{\mu_1}\cdots k_{\mu_N}}
{(k^2)^{N+2}(k+q)^2}
\end{eqnarray}
which is again of the self energy type.
\section{Mellin Barnes Techniques}
A method which turns out to be very useful to compute three-point and
four-point functions even up to two-loop level is given by the Mellin-Barnes
technique \cite{usda}. A simple Mellin-Barnes transformation takes the form
\begin{eqnarray}
\label{eqn3.1}
(A_1+A_2)^{-\nu} = \frac{1}{2\pi i} \int_{-i\infty}^{i\infty} A_1^{\sigma}
A_2^{-\nu-\sigma} \frac{\Gamma(-\sigma)\Gamma(\nu+\sigma)}{\Gamma(\nu)}\,,
\end{eqnarray}
with $|arg(A_1)-arg(A_2)|<\pi$ and the poles in $\sigma$ are located on the 
real axis.  To illustrate this technique we take as an example
the three-point function where the propagators are raised to an arbitrary power
\begin{eqnarray}
\label{eqn3.2}
J(n;\nu_1,\nu_2,\nu_3)&=&\int d^n k \frac{1}{[(q_1+k)^2-m_1^2]^{\nu_1}
[(q_2+k)^2-m_2^2]^{\nu_2}[(q_3+k)^2-m_3^2]^{\nu_3}}
\nonumber\\[2ex]
&=& i^{1-n} \pi^{n/2} \frac {\Gamma(\sum_i \nu_i-n/2)}{\prod_i \Gamma(\nu_i)}
\int \prod_i d \alpha_i \, \alpha_i^{\nu_i-1} \,\delta(1- \sum_j \alpha_j)
\nonumber\\[2ex]
&& \times \frac{1}{[-\sum_j \alpha_j m_j^2 + \alpha_2\alpha_3 p_1^2
+\alpha_1\alpha_3 p_2^2+\alpha_1\alpha_2 p_3^2]^{\sum_k \nu_k - n/2}}\,,
\end{eqnarray}
where $p_1=q_3-q_2$, $p_2=q_1-q_3$ and $p_3=q_2-q_1$. Next we apply the
double Mellin Barnes integral transformation given by
\begin{eqnarray}
\label{eqn3.3}
\frac{1}{(X+Y+Z)^a}&=&\frac{1}{Z^a}  
\int_{-i\infty}^{i\infty} \frac{du}{2\pi i} \int_{-i\infty}^{i\infty} 
\frac{dv}{2\pi i} \, 
\frac{\Gamma(a+u+v)\Gamma(-u)\Gamma(-v)} {\Gamma(a)} 
\nonumber\\[2ex]
&& \left (\frac{X}{Z} \right )^u\left (\frac{Y}{Z} \right )^v\,,
\end{eqnarray}
so that the integral in Eq. (\ref{eqn3.2}) can be written as
\begin{eqnarray}
\label{eqn3.4}
&&J(n;\nu_1,\nu_2,\nu_3)=\frac{\pi^{n/2}i^{1-n}}{\prod_j \Gamma(\nu_j)}
\int \prod_i d \alpha_i \, \alpha_i^{\nu_i-1} \,\delta(1- \sum_j \alpha_j)
\left ( \alpha_1\alpha_2 p_3^2 \right )^{n/2-\sum_k \nu_k} 
\nonumber\\[2ex]
&& \int_{-i\infty}^{i\infty} \frac{du}{2\pi i} \int_{-i\infty}^{i\infty} 
\frac{dv}{2\pi i}
\Gamma(\sum_i \nu_i - n/2 + u + v) \Gamma(-u)\Gamma(-v) 
\left (\frac{\alpha_3~p_1^2}{\alpha_1~p_3^2} \right )^u 
\left (\frac{\alpha_3~p_2^2}{\alpha_2~p_3^2} \right )^v \,,
\end{eqnarray}
where we have put for simplicity $m_i^2=0$. Using
\begin{eqnarray}
\label{eqn3.5}
\prod_i \int_0^1 d\alpha_i \delta(1-\sum_j \alpha_j) \alpha_1^a\alpha_2^b
\alpha_3^c = \frac{\Gamma(1+a)\Gamma(1+b)\Gamma(1+c)}{\Gamma(1+a+b+c)}\,,
\end{eqnarray}
the three-point function reads
\begin{eqnarray}
\label{eqn3.6}
&& J(n;\nu_1,\nu_2,\nu_3)= 
\nonumber\\[2ex]
&&\pi^{n/2} i^{1-n} \int_{-i\infty}^{i\infty} \frac{du}{2\pi i} 
\int_{-i\infty}^{i\infty} \frac{dv}{2\pi i}
\frac{\Gamma(n/2-\nu_1-\nu_3)\Gamma(n/2-\nu_2-\nu_3)\Gamma(\nu_3+u+v)}
{\prod_i \Gamma(\nu_i)\Gamma(n- \sum_j \nu_j)}
\nonumber\\[2ex]
&& \Gamma(\sum_k \nu_k +u+v-n/2)\Gamma(-u)\Gamma(-v) 
\left (\frac{p_1^2}{p_3} \right )^u \left (\frac{p_2^2}{p_3} \right )^u
\left (p_3 \right )^{n/2-\sum_j \nu_j}\,.
\end{eqnarray}
A special case is $\nu_1+\nu_2+\nu_3=n$ (Uniqueness Condition). After
application of the residue theorem to expression (\ref{eqn3.6}) one obtains
\begin{eqnarray}
\label{eqn3.7}
J(n;\nu_1,\nu_2,\nu_3)=\pi^{n/2} i^{1-n} \prod_{j=1}^3 
\frac{\Gamma(n/2-\nu_j)}{\Gamma(\nu_j)} \left (p_j^2 \right )^{\nu_j-n/2}\,.
\end{eqnarray}
\begin{figure}
\begin{center}
  \begin{picture}(180,70)(0,0)
  \ArrowLine(10,35)(0,35)
  \Line(10,35)(60,60)
  \Line(10,35)(60,10)
  \ArrowLine(60,60)(80,70)
  \ArrowLine(60,10)(80,0)
  \Line(35,48)(35,22)
  \Line(60,60)(60,10)
  \Text(5,32)[t]{$p_3$}
  \Text(70,60)[t]{$p_1$}
  \Text(70,17)[t]{$p_2$}

  \ArrowLine(110,35)(100,35)
  \Line(110,35)(160,60)
  \ArrowLine(160,60)(180,70)
  \Line(110,35)(160,10)
  \ArrowLine(160,10)(180,0)
  \Line(135,48)(160,10)
  \Line(160,60)(135,22)
  \Text(105,32)[t]{$p_3$}
  \Text(170,60)[t]{$p_1$}
  \Text(170,17)[t]{$p_2$}

  \end{picture}
  \caption[]{Two-loop triangle graphs.}
  \label{fig5}
\end{center}
\end{figure}
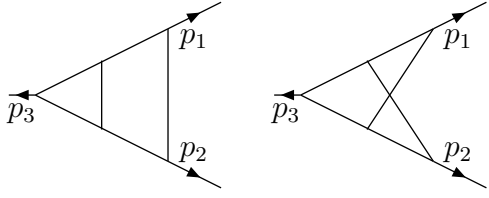
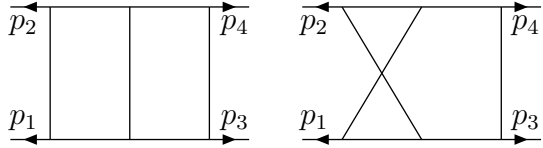
\begin{figure}
\begin{center}
  \begin{picture}(200,70)(0,0)
  \ArrowLine(15,60)(0,60)
  \ArrowLine(75,60)(90,60)
  \ArrowLine(15,10)(0,10)
  \ArrowLine(75,10)(90,10)
  \Line(15,60)(75,60)
  \Line(15,10)(75,10)
  \Line(15,60)(15,10)
  \Line(75,60)(75,10)
  \Line(45,60)(45,10)

  \Text(5,21)[t]{$p_1$}
  \Text(5,57)[t]{$p_2$}
  \Text(85,21)[t]{$p_3$}
  \Text(85,57)[t]{$p_4$}

  \ArrowLine(125,60)(110,60)
  \ArrowLine(185,60)(200,60)
  \ArrowLine(125,10)(110,10)
  \ArrowLine(185,10)(200,10)
  \Line(125,60)(185,60)
  \Line(125,10)(185,10)
  \Line(125,60)(155,10)
  \Line(155,60)(125,10)
  \Line(185,60)(185,10)
  \Text(115,21)[t]{$p_1$}
  \Text(115,57)[t]{$p_2$}
  \Text(195,21)[t]{$p_3$}
  \Text(195,57)[t]{$p_4$}

  \end{picture}
  \caption[]{Two-loop box graphs.}
  \label{fig6}
\end{center}
\end{figure}
The Mellin Barnes techniques has been successfully applied to compute
two-loop graphs. Examples are the three-point functions in Fig. \ref{fig5}
and the four point functions in Fig. \ref{fig6} which are computed in 
\cite{usda} for $p_i^2 \not =0$. Recent progress has been made for the two-loop
box graphs in Fig. \ref{fig6} which has been computed for all external 
momenta on-shell ($p_i^2 =0$) \cite{sm}, \cite{ta}. This is very useful for
applications to the second order corrections in Bhabha scattering and in di-jet
production.
\section{Negative Dimension Approach}
This method has been recently developed in \cite{hari} \cite{angl} 
to express one-loop integrals containing different masses into special 
functions. To illustrate the technique we take the two-point function
as an example. The latter is given by the integral
\begin{eqnarray}
\label{eqn4.1}
I_2^n(\nu_1,\nu_2,q^2,M_1^2,M_2^2)= \int \frac{d^nk}{i\pi^{n/2}} 
\frac{1}{A_1^{\nu_1} A_2^{\nu_2}}\,,
\end{eqnarray}
Using the Schwinger representation for the propagator
\begin{eqnarray}
\label{eqn4.2}
\frac{1}{A_i^{\nu_i}}=\frac{(-1)^{\nu_i}}{\Gamma(\nu_i)} \int_0^{\infty}
d x_i \, x_i^{\nu_i-1} \, exp \left(x_iA_i \right )\,,
\end{eqnarray}
expression (\ref{eqn4.1}) becomes equal to
\begin{eqnarray}
\label{eqn4.3}
I_2^n=\prod_{i=1}^2 \frac{(-1)^{\nu_i}}{\Gamma(\nu_i)}\int_0^{\infty}
d x_i \, x_i^{\nu_i-1} \int
\frac{d^nk}{i\pi^{n/2}}exp \left (\sum_{j=1}^2 x_jA_j \right )\,.
\end{eqnarray}
The expression above can be evaluated in two different ways. The first way
proceeds by expanding the exponents in a power series. This yields
\begin{eqnarray}
\label{eqn4.4}
I_2^n&=&\prod_{i=1}^2 \frac{(-1)^{\nu_i}}{\Gamma(\nu_i)}
\int_0^{\infty} d x_i \, x_i^{\nu_i-1}
\int \frac{d^nk}{i\pi^{n/2}} \sum_{n_1=0}^{\infty} \frac{(x_1A_1)^{n_1}}{n_1!}
\sum_{n_2=0}^{\infty}\frac{(x_2A_2)^{n_2}}{n_2!}=
\nonumber\\[2ex]
&& \prod_{i=1}^2 \frac{(-1)^{\nu_i}}{\Gamma(\nu_i)}
\int_0^{\infty} d x_i \, x_i^{\nu_i-1}
\sum_{n_1=0}^{\infty}\sum_{n_2=0}^{\infty} I_2^n(-n_1,-n_2,q^2,M_1^2,M_2^2)
\prod_{j=1}^2 \frac{x_j^{n_j}}{n_j!}\,.
\end{eqnarray}
Using the identity
\begin{eqnarray}
\label{eqn4.5}
\frac{(-1)^{\nu_i}}{\Gamma(\nu_i)}\int_0^{\infty} d x_i \, x_i^{\nu_i+n_j-1}=
\delta_{\nu_i+n_j,0}\,,
\end{eqnarray}
we obtain
\begin{eqnarray}
\label{eqn4.6}
I_2^n=I_2^n(\nu_1,\nu_2,q^2,M_1^2,M_2^2)\prod_{i=1}^2 \frac{1}{\Gamma(1-\nu_i)}
\,.
\end{eqnarray}
There is a second way to compute Eq. (\ref{eqn4.3}). First one shifts the
momentum $k$ which becomes
$k=k'- q~x_2/(x_1+x_2)$. Subsequently one uses the following identities
\begin{eqnarray}
\label{eqn4.7}
\int \frac{d^nk'}{i\pi^{n/2}} \left ( k'^2 \right )^l=l!\,\delta_{l+n/2,0}
\qquad \int \frac{d^nk'}{i\pi^{n/2}}  exp \left( \alpha k'^2 \right ) =
\frac{1}{\alpha^{n/2}}\,.
\end{eqnarray}
Notice that since $l>0$ one obtains $n<0$. This the reason for the name 
"Negative Dimension Approach". The expression in Eq. (\ref{eqn4.3}) can now
be written as
\begin{eqnarray}
\label{eqn4.8}
I_2^n=\prod_{i=1}^2 \frac{(-1)^{\nu_i}}{\Gamma(\nu_i)}
\int_0^{\infty} d x_i \, x_i^{\nu_i-1}
\frac{1}{(x_1+x_2)^{n/2}} exp \left (\frac{x_1x_2}{x_1+x_2} q^2 \right )
exp \left ( -\sum_{j=1}^2 x_jM_j^2 \right )\,.
\end{eqnarray}
The integrand can be written as
\begin{eqnarray}
\label{eqn4.9}
&& \frac{1}{(x_1+x_2)^{n/2}} exp \left (\frac{x_1x_2}{x_1+x_2} q^2 \right )
exp \left ( -\sum_{j=1}^2 x_jM_j^2 \right )
\nonumber\\[2ex]
&& = \sum_{l=0}^{\infty} \frac{1}{l!} (x_1+x_2)^{-l-n/2}(x_1x_2q^2)^l
 \sum_{m=0}^{\infty} \frac{(-x_1M_1^2-x_2M_2^2)^m}{m!}\,.
\end{eqnarray}
Using the identities
\begin{eqnarray}
\label{eqn4.10}
(\sum_{i=0}^M x_iA_i)^N= \Gamma(N+1) \prod_{k=1}^M 
\sum_{i_k=0}^{\infty}\frac{x_{i_k}A_{i_k}} {i_k!} 
\quad \mbox{with} \quad \sum_{k=1}^M i_k=N\,,
\end{eqnarray}
one can write
\begin{eqnarray}
\label{eqn4.11}
I_2^n&=&\prod_{i=1}^2 \frac{(-1)^{\nu_i}}{\Gamma(\nu_i)}
\int_0^{\infty} d x_i \, x_i^{\nu_i-1}
\sum_{l=0}^{\infty} \frac{(x_1x_2q^2)^l}{l!} \Gamma(1-l-n/2) 
\sum_{p_1=0}^{\infty} \sum_{p_2=0}^{\infty} \frac{x_1^{p_1}}{p_1!}
\frac{x_2^{p_2}}{p_2!} 
\nonumber\\[2ex]
&& \sum_{m_1=0}^{\infty} \sum_{m_2=0}^{\infty}
\frac{(-x_1M_1^2)^{m_1}}{m_1!} \frac{(-x_2M_2^2)^{m_2}}{m_2!}
\quad \mbox{with} \quad p_1+p_2=-l-n/2\,.
\end{eqnarray}
Equating Eqs. (\ref{eqn4.6}) and  (\ref{eqn4.11}) one obtains the result
\begin{eqnarray}
\label{eqn4.12}
I_2^n(\nu_1,\nu_2,q^2,M_1^2,M_2^2)&=& \sum_{l,p_i,m_i} \frac{\Gamma(1-l-n/2)
\Gamma(1-\nu_1)\Gamma(1-\nu_2)}
{\Gamma(1+l)\Gamma(1+p_1)\Gamma(1+p_2)\Gamma(1+m_1)\Gamma(1+m_2)}
\nonumber\\[2ex]
&& (q^2)^{l+m_1+m_2} \times \left (\frac{-M_1^2}{q^2} \right )^{m_1} 
\left (\frac{-M_2^2}{q^2} \right )^{m_2}\,,
\end{eqnarray}
with
\begin{eqnarray}
\label{eqn4.13}
p_1+p_2=-l-n/2 \quad p_1=l-\nu_1-m_1 \quad p_2=l-\nu_2-m_2\,.
\end{eqnarray}
Choosing two independent summation indices out of five i.e. $l,p_1,p_2,m_1,m_2$
one can express $I_2^n$ into the following special functions \cite{angl}.
given by the Hyper-geometric Functions $_2F_1$, $_3F_2$, the Appell 
Functions $F_i$ ($i=1-4$), the Horn Function $H_2$ and the Kamp\'e de F\'eriet 
Functions $S_i$ ($i=1,2$).
In the case of the two-point function in Eq. (\ref{eqn4.1}) there are
eight possibilties to express the integral into the special functions
above depending on the convergence of the sums in a specific kinematic 
region. If we choose $|M_i^2/q^2|<1$ one has to take $m_1,m_2$. Using the 
following identity
\begin{eqnarray}
\label{eqn4.14}
\frac{\Gamma(z-n)}{\Gamma(z)}=(-1)^n \frac{\Gamma(1-z)}{\Gamma(1-z+n)}\,,
\end{eqnarray}
the expression in Eq. \ref{eqn4.12} can be written as
\begin{eqnarray}
\label{eqn4.15}
&&I_2^n(\nu_1,\nu_2,q^2,M_1^2,M_2^2)=\sum_{m_1=0}^{\infty}\sum_{m_2=0}^{\infty} 
\frac{\Gamma(1-n+\nu_1+\nu_2+m_1+m_2)}{\Gamma(n/2+1-\nu_1-\nu_2-m_1-m_2)}
\nonumber\\[2ex]
&&\times \frac{\Gamma(1-\nu_1) \Gamma(1-\nu_2)}{\Gamma(1+m_1)\Gamma(1+m_2)}
(q^2)^{n/2-\nu_1-\nu_2} \times \left (\frac{-M_1^2}{q^2} \right )^{m_1}
\left (\frac{-M_2^2}{q^2} \right )^{m_2} 
\nonumber\\[2ex]
&&=(-1)^{n/2} (q^2)^{n/2-\nu_1-\nu_2} \frac{\Gamma(n/2-\nu_1)\Gamma(n/2-\nu_2)
\Gamma(\nu_1+\nu_2-n/2)}{\Gamma(\nu_1)\Gamma(\nu_2)\Gamma(n-\nu_1-\nu_2)}
\nonumber\\[2ex]
&& F_4(1+\nu_1+\nu_2-n,\nu_1+\nu_2-n/2,1+\nu_1-n/2,1+\nu_2-n/2,\frac{M_1^2}{q^2}
,\frac{M_2^2}{q^2})\,.
\end{eqnarray}
Here the fourth Appell Function is given by
\begin{eqnarray}
\label{eqn4.16}
F_4(\alpha,\beta,\gamma,\gamma',x,y)&=&\sum_{m,n=0}^{\infty}
\frac{\Gamma(\alpha+m+n)}{\Gamma(\alpha)}
\frac{\Gamma(\beta+m+n)}{\Gamma(\beta)}\frac{\Gamma(\gamma)}{\Gamma(\gamma+m)}
\frac{\Gamma(\gamma')}{\Gamma(\gamma'+n)}
\nonumber\\[2ex]
&& \times \frac{x^m}{\Gamma(m+1)} \frac{y^n}{\Gamma(n+1)}\,.
\end{eqnarray}
We conclude that using the method above one can find different representations
for an one-loop integral in different kinematic regions depending on the
radius of convergence of the ratios between the scales $q^2,M_1^2,M_2^2$. It is 
unclear whether 
this method, which seems to us very complicate, can be used to compute 
two-loop integrals.


\begin{thebibliography}{99}
\bibitem{zn}
W.L. van Neerven, Nucl. Phys. {\bf B268} (1986) 453;\\
E.B. Zijlstra and W.L. van Neerven, Nucl. Phys. {\bf B383} (1992) 525.
\bibitem{move}
S. Moch and J.A.M. Vermaseren, hep-ph/9912355, hep-ph/9909269.
\bibitem{chtk}
K.G. Chetyrkin and F.V. Tkachov, Nucl. Phys. {\bf B192} (1981) 159. 
\bibitem{lari}
S.A. Larin, T. van Ritbergen and J.A.M. Vermaseren, Nucl. Phys. {\bf B427} 
(1994) 41;\\
S.A. Larin, P. Nogueira, T. van Ritbergen and J.A.M. Vermaseren, Nucl. Phys. 
{\bf B492} 338.
\bibitem{usda}
N.I. Ussyukina and A.I. Davydychev, Phys. Lett. {\bf B298} (1993) 363,
ibid. {\bf B332} (1994) 159, {\bf B348} (1993) 503.
\bibitem{sm}
V.A. Smirnov, Phys.Lett. {\bf B460} (1999) 397.
\bibitem{ta}
J.B. Tausk, Phys.Lett. {\bf B469} (1999) 225. 
\bibitem{hari}
L.G.Halliday and R.M. Ricotta, Phys.Lett. {\bf B193} (1987) 241;\\
A.T. Suzuki and A.G.M. Schmidt, Eur. Phys. J. {\bf C5} (1998) 175, Phys. Rev.
{\bf D58} (1998) 047701.
\bibitem{angl}
C. Anastasiou, E.W.N. Glover and C. Oleari, hep-ph/9907494.
\end{thebibliography}
\end{document}